# A Complex Network-Based Framework for Revealing Spatiotemporal Rainfall Structures

Ajay Bankar[1, 2*], Praveenkumar Venkatesan[1, 2], Rakesh V[3], Gaurav Chopra[4], and R I Sujith[1, 2]

[1] Department of Aerospace Engineering, Indian Institute of Technology (IIT) Madras, Chennai–600036, India
[2] Centre of Excellence for Studying Critical Transitions in Complex Systems, IIT Madras, Chennai–600036, India
[3] CSIR National Institute of Data Science and Artificial Intelligence, Bengaluru–560037, India
[4] Department of Applied Mechanics, IIT Delhi, New Delhi–110016, India
* corresponding author: ajaybankar123@gmail.com

## Abstract

Reliable spatio-temporally continuous precipitation estimates are essential for hydrological and climate studies, yet satellite-derived products show varying skill across different regions. Conventional evaluations against ground observations often reveal only modest differences among datasets, limiting their ability to clearly discriminate product quality. Here, using complex network theory, we examine spatial variability of summer monsoon rainfall over a topologically complicated region Karnataka, India with six widely used satellite-derived rainfall datasets. The network-based analysis reveals substantial contrasts among the satellite products in representation of rainfall extremes and teleconnection pattern. Among all satellite products, GSMaP-ISRO is the most consistent with the ground-based rainfall observations, effectively distinguishing large-scale coherence from localized variability, likely due to bias correction using IMD gridded dataset. The network-based method offers a discriminative framework than traditional metrics and highlight the need for improved representation of localized extremes in monsoon regions in complex terrain.

## Plain Language Summary

Satellite-based rainfall datasets provide valuable information over regions where ground observations are sparse, but their accuracy and representation of rainfall patterns can vary due to differences in retrieval methods and limited validation data. In this study, we assess six widely used precipitation products over Karnataka, India, a region with complex terrain and strong spatial west-east rainfall gradient, by comparing them with dense rain-gauge observations. Instead of relying on traditional error statistics, we use a network-based approach to examine how rainfall is organized in space and time. This analysis reveals clear differences among datasets in their ability to represent regional rainfall structure, particularly over interior and topographically complex areas. The results show that some products more closely reproduce the observed rainfall organization, while others exhibit overly localized or overly coherent patterns. Overall, the study demonstrates that network-based methods provide additional and physically meaningful insight into the performance of satellite precipitation products, especially for understanding regional rainfall variability and extremes.

## 1. Introduction

Accurate estimation of precipitation is fundamental to understanding the Earth's hydrological and energy cycles, yet it remains one of the most challenging meteorological variable to observe and quantify due to its high spatial and temporal variability (Beck et al., 2019). Reliable precipitation observations are essential in a wide range of applications, including hydrology, hydroelectric power generation, food security, as well as studying climate trends and variability (Prakash et al., 2018). Ground-based observations from rain gauges and weather radars serve as the primary sources of precise precipitation data (Li et al., 2021); however, these measurements are often limited by spatial and temporal gaps.

Advancements in remote sensing have enabled satellite-derived rainfall estimates to provide broader spatial coverage and higher temporal sampling frequencies (Hong et al., 2004), making them attractive alternatives. To better understand the ecological and environmental variability of a region, accurate and spatially consistent precipitation data are essential (Hu et al., 2016). As a result, a wide range of gridded precipitation products has been developed, differing in purpose, spatial and temporal resolution, and methodological approach (Beck et al., 2017). However, inconsistencies remain among these products due to algorithmic assumptions and limitations in the underlying observational sources (Sun et al., 2018). Furthermore, the decline in rain-gauge coverage globally has intensified reliance on satellite and reanalysis-based datasets (Mankin et al., 2025). Blended precipitation products aim to mitigate these limitations by merging high-accuracy passive microwave (PMW) estimates from low Earth orbit (LEO) satellites with high-frequency infrared (IR) estimates from geostationary satellites (Tapiador et al., 2004). While this merging improves spatiotemporal resolution, satellite-derived estimates remain affected by sensor limitations, retrieval uncertainties, and terrain complexity (AghaKouchak et al., 2012). Therefore, ground-based observations remain indispensable for calibrating satellite products. Several studies have shown that satellite datasets adjusted using daily gauge observations perform better than those using coarser monthly data (Kumar et al., 2022; Beck et al., 2017).

Validating these satellite-derived precipitation datasets is essential before using them in climate modeling, hydrological simulation, or water resource planning. Numerous global and regional studies have evaluated their accuracy (Boers et al., 2015; Khandu et al., 2016; Beck et al., 2017; Sun et al., 2018; Maggioni et al., 2016; Awange et al., 2016; Chen et al., 2020; Macharia et al., 2022; Li et al., 2021; Saikrishna et al., 2021; Prakash et al., 2018; Kumar et al., 2022;). For instance, Prakash et al. (2018) reported considerable errors in GPM-based products over orographic regions. Sun et al. (2018) reviewed 30 global datasets and highlighted regional disparities, with especially high uncertainty over tropical oceans, mountainous terrains, and high-latitude areas. Li et al. (2020) further showed that IMERG has limited ability to detect light rainfall events (<5 mm/day), particularly over deserts and mountains.

Most of the studies involving dataset validations rely on bias, RMSE, and correlation which treats errors in scalar form. Though these error metric are useful they ignore spatial coherence, co-occurrence of events and topology. Complex network metrics (Boers et al., 2015) capture spatiotemporal connectivity and clustering. Also nonlinear measure event synchronization quantifies extreme rainfall events co-occurrence. Using complex network metrics we show that widely used rainfall products systematically distort spatial dependence and extreme event-synchronization, revealing structural biases invisible by standard metrics.

Despite growing efforts to validate precipitation products, many previous studies have relied on sparse rain-gauge networks or have been limited to inter-product comparisons, which constrain their ability to evaluate accuracy under varying climatic and topographic conditions. In contrast, this study evaluates six precipitation datasets—CHIRPS, PERSIANN-CCS, MSWEP, IMERG, GSMaP_ISRO, and AgERA5—against a dense network of telemetric rain gauges across Karnataka, India, a region with diverse climate zones and complex terrain. Although some validation studies have been conducted over portions of this region, they either examined only a few precipitation products or were confined to the Western Ghats (WG), without utilizing the comprehensive rain-gauge network employed in the present study.

Overall, this study not only benchmarks the performance of widely used precipitation products over a climatically sensitive region but also demonstrates the added value of network-based techniques in capturing rainfall connectivity—an important feature for understanding extreme rainfall dynamics and improving early warning systems.

The remaining parts of the paper are organized as following: Section 2 describes study area, methodology; Section 3 discusses findings of the study; and conclusions are summarized in Section 4.

## 2. Study Area and Methods

### 2.1. Study Area

Karnataka (S. Figure 1) is a state in southern India between 11.5ºN–18.5ºN latitudes and 74ºE–78.5ºE longitudes, lies along the western edge of Deccan Peninsula. It exhibits significant spatial and temporal variability in rainfall due to its diverse topography. The state receives 74% of its annual rainfall during the southwest monsoon (SWM) season, spanning from June to September (JJAS). Rainfall distribution varies significantly from west to east, with coastal districts receiving over 4500 mm of rainfall annually, while the eastern interior regions receive as little as 540 mm (Karnataka Annual Weather Report 2023). This gradient gives rise to varied climatic zones, ranging from humid to semi-humid conditions in the coastal and WG regions to semi-arid to arid climates in the interior plateau.

Based on climatic and topographic characteristics, Karnataka is categorized into four major subregions: the coastal region, Malnad (part of WG mountain system), south interior Karnataka (SIK), and north interior Karnataka (NIK). The coastal belt, a narrow strip along the Arabian Sea on the windward side of WG, receives intense monsoonal rainfall averaging around 3518 mm annually, driven by orographic lifting. Adjacent to the coast, the Malnad region–formed by the Sahyadri range of the WG–is a mountainous, forest-covered zone with annual rainfall around 1950 mm, exhibiting a humid to semi-humid climate. Further east, the SIK region is a semi-arid plateau with elevations ranging from 600 to 900 m, surrounded by the WG on the west and south, and receiving an annual rainfall of 714 mm. The NIK region lies further northeast, characterized by a drier plateau landscape, part of the Deccan Trap, with elevations between 300 and 600 m and an average annual rainfall of about 702 mm. Overall, Karnataka exhibits strong west-to-east rainfall gradients, with the state average annual rainfall around 1151 mm, predominantly influenced by the SWM. This strong spatial contrast, shaped by orographic effects and mesoscale systems, makes Karnataka an ideal region to study the performance of precipitation datasets across varying hydroclimatic conditions.

## 2.2 Methodology

Due to differences in spatial and temporal resolutions across the various precipitation products, it was necessary to standardize all datasets to a common spatial and temporal resolution for meaningful comparison. Except for GSMaP_ISRO, all other precipitation products are natively available in daily format. Therefore, the hourly GSMaP_ISRO data was first aggregated to a daily timescale. To facilitate point-to-point comparisons, all gridded precipitation datasets were interpolated to the rain-gauge station locations using the Inverse Distance Weighting (IDW; Shepard 1968) method. All precipitation products are available in the 0000 UTC to 2400 UTC time window, whereas the KSNDMC rain-gauge observations are recorded at 15-minute intervals. These station observations were accordingly aggregated to daily values consistent with the UTC time format to ensure accurate temporal alignment during the comparison. The analysis focuses on the Indian summer monsoon season (JJAS) for a 12-year period (2011–2022). However, as the MSWEP dataset is only available up to the year 2020, its analysis was limited to that period.

To examine the spatial structure and similarity of rainfall patterns across datasets, we employed methods from complex network theory (Stolbova et al., 2014; Boers et al., 2014; Boers et al., 2015). In this framework, a network (or graph) is defined as a collection of nodes (vertices) and links (edges) that represent relationships between nodes. Each station location represents a node associated with its corresponding daily rainfall time series. A link between two nodes is established if they exhibit a statistically significant relationship. A correlation-based rainfall network was constructed by computing the Pearson correlation coefficient ($r$) between daily rainfall time series at all pairs of stations. Significance test was performed, and only correlations with p-value less than 0.05 were retained. To ensure meaningful interconnectivity, two nodes $i$ and $j$ were connected only if their correlation $C_{ij}$ exceeds the 95$^{th}$ percentile threshold ($\gamma$), thereby preserving the strongest correlations. These significant correlations were represented using a binary adjacency matrix $A$, where each element $A_{ij}$, was assigned a value of 1 if there exists a link between node $i$ and $j$, and 0 otherwise. This formulation results in an unweighted and undirected network, such that $A_{ij}$= $A_{ji}$, making the adjacency matrix symmetric. The degree of a node, defined as the number of connections it possesses (Newman, 2018) was computed by summing the links in each row of the adjacency matrix. In this context, the degree represents the number of strong correlations that a station shares with other stations. The ($i$, $j$)$^{th}$ element of the adjacency matrix and degree ($k_i$) of the node $i$ are defined as follows:

$$A_{ij} = \begin{cases} 1 & if\, r > \gamma \\ 0 & if\, r < \gamma \end{cases} \tag{5}$$

$$k_i = \sum_{j=1}^{n} A_{ij} \tag{6}$$

where, $n$ is the total number of nodes.

To assess the similarity between precipitation products and observed rainfall, we also computed the Salton cosine similarity (Salton and McGill, 1983) between the corresponding nodes. For a given node $i$, it is defined as:

$$S_i = \frac{K_i \cdot K_i'}{\sqrt{k_i k_i'}} \tag{7}$$

where $K_i$ and $K_i'$ are the $i^{th}$ row vectors of adjacency matrices of the networks corresponding to the observed rainfall and the gridded rainfall product, respectively, and $k_i$ and $k_i'$ represent the associated node degrees.

To explore the temporal synchronization of extreme rainfall events (EREs), we employed a nonlinear measure known as event synchronization (ES; Boers et al., 2014). EREs at each node were defined as days when daily rainfall exceeded the 95$^{th}$ percentile threshold of wet days (days having rainfall more than 2.5 mm). For each node, an event series was then constructed by retaining only days above this threshold from the daily rainfall time series. An ES-based network was constructed by identifying co-occurring extreme events within a maximum allowed lag ($\tau_{max}$) of 5 days. Events occurring at different spatial locations were considered synchronized if their timing fell within this $\tau_{max}$ window. This procedure gives a symmetric strength matrix Q. To test statistical significance, 1000 surrogate event series were generated by randomly permuting the events, thereby ensuring synchronization did not arise by chance but reflected genuine physical processes. Only those strength values exceeding the significant value (> 95%) were retained. Further methodological details on ES can be found in (Venkatesan et al., 2025).

## 3. Results

### 3.1. Spatial Distribution of Rainfall Errors

Spatial distribution of errors in daily mean rainfall across Karnataka is shown in Figure 1. Pronounced differences among the datasets emerge over the high-rainfall regions of the coastal belt and the WG (S. Figure 2). In particular, PERSIANN-CCS exhibits a strong negative bias (Figure 1a) on the windward slopes of the WG, substantially underestimating rainfall. This deficiency is primarily attributed to its reliance on cloud-top brightness temperature, which limits its ability to capture warm rain processes that dominate in this orographically forced environment. In contrast, the remaining datasets tend to overestimate rainfall over this complex terrain by up to ~20 mm/day. Across the northern districts of the NIK region, all products consistently overestimate rainfall by approximately 5–10 mm/day. This systematic bias likely reflects the complex interaction between monsoonal flow and regional topography, where mesoscale convection and localized moisture convergence are amplified in gridded precipitation products. Over the SIK region, the datasets perform comparatively well, with biases generally within 5 mm/day. RMSE (Figure 1g–l) exhibits the largest values over the WG and coastal regions, consistent with the elevated biases in these areas, which receive the highest seasonal rainfall during the monsoon. These large errors underscore the difficulty of accurately representing intense orographic precipitation, multiscale interactions, and complex cloud microphysics in satellite-based and reanalysis products. Outside the WG and coastal belt, differences among the datasets are relatively modest, highlighting the limitations of conventional scalar metrics in clearly discriminating dataset performance.

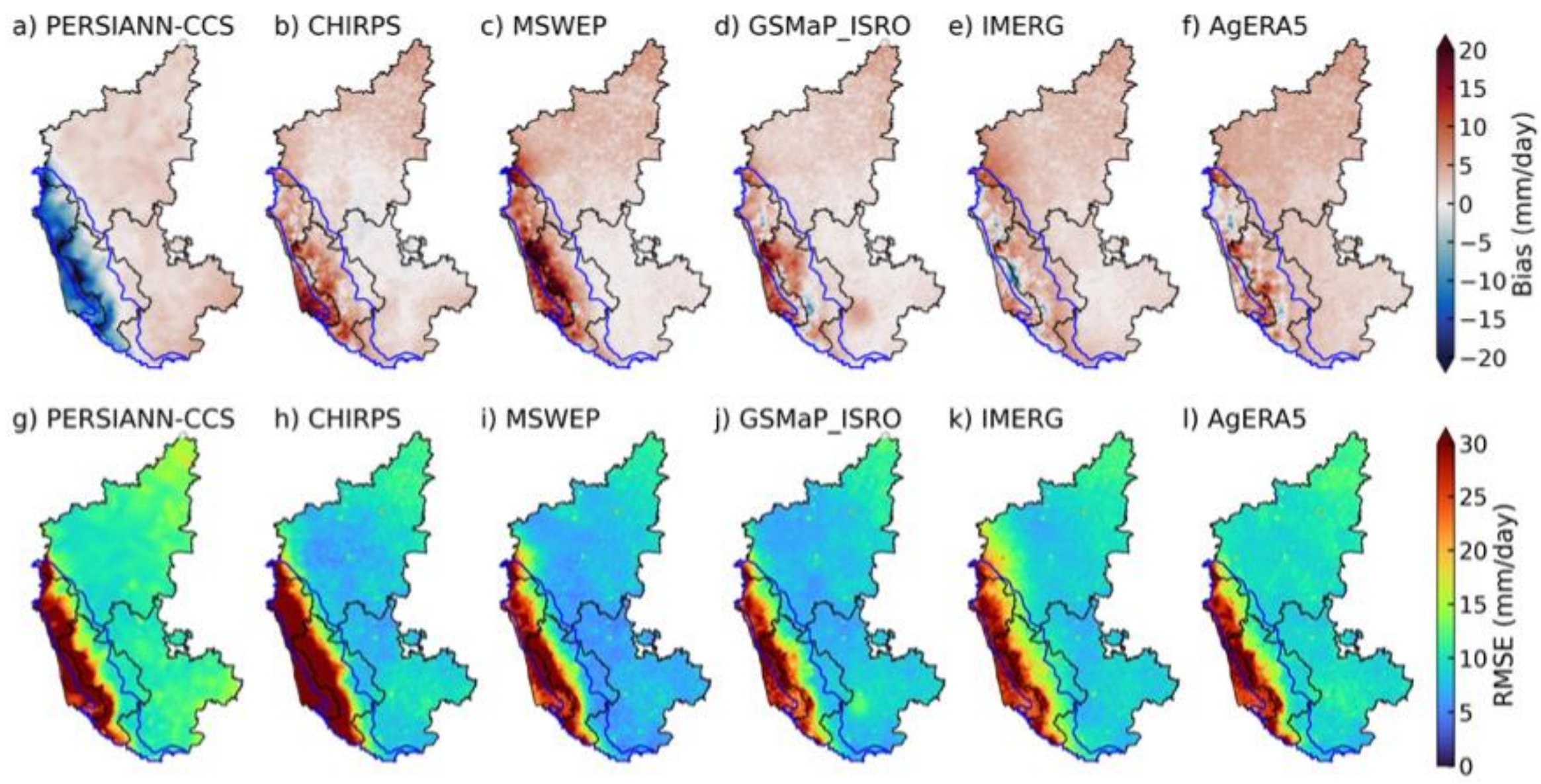

**Figure 1.** Spatial distribution of daily mean bias (a–f) and RMSE (g–l) in monsoon (JJAS) rainfall averaged over 2011–2022 (2011–2020 for MSWEP).

### 3.2. Rainfall Correlation Network Analysis

Spatial distribution of the degree in the rainfall correlation network constructed from daily mean rainfall during the summer monsoon season is presented in Figure 2. The degree here represents the number of other stations with which a given station shares a significant rainfall correlation (above the 95$^{th}$ percentile), providing insights into the spatial coherence of rainfall patterns across Karnataka. The analysis reveals that the WG and central parts of SIK exhibit the highest degree values. This indicates that rainfall in these regions is strongly correlated with many other parts of the state. The PDF of the link distance (Figure 2h1 and h2) shows a sharp peak at short distances with a long tail extending beyond 300-km, especially in the coastal and Malnad regions, indicating both local and long-range rainfall connections due to organized systems. In Malnad region, broad and flat PDF curve of node degree (Figure 2i2) shows a well distributed range of connectivity. Such high degree regions are typically associated with large-scale, organized rainfall systems, including monsoon depressions, low-pressure systems, and persistent moisture influx from the Arabian sea. In particular, orographic lifting along the WG enhances vertical motion, further amplifying rainfall intensity and spatial coverage. These physical processes result in widespread and simultaneous rainfall, reflecting high spatial coherence. These regions effectively act as rainfall hubs, areas that anchor broad-scale monsoon activity. In contrast, the NIK region and the northern and eastern parts of SIK show lower degrees. This implies that rainfall in these areas is strongly correlated with only a limited number of other stations. The reduced connectivity suggests that rainfall in these regions is governed by localized convective events, or rain shadow effects due to their leeward position relative to WG. Rainfall in these areas tend to be more isolated, patchy, and intermittent, leading to weaker spatial coherence.

When extended to gridded products, the rainfall correlation networks reveal pronounced differences in coherence structure among the gridded products. GSMaP_ISRO most closely reproduces the observed pattern, capturing high-degree zones over the WG and central SIK, consistent with large-scale monsoon forcing. This agreement is supported by degree PDF with multimodal behavior indicating high connectivity pattern. However, in SIK and NIK, it exhibits a more structured grid pattern, with elevated degrees along linear bands and reduced connectivity within gridded cells. This artifact likely stems from the interpolation and smoothing process involved in producing gridded data. PERSIANN-CCS reproduces high connectivity in central SIK, with degree and link-distance distributions matching observations, but fails over the WG and coastal Karnataka, with link-distance distributions confined to less than 100-km, consistent with its limited skill in complex terrain; it also overestimates coherence in NIK. CHIRPS shows strongly localized connections over the WG and coastal region, and unrealistically high degrees in NIK. MSWEP captures the main coherence features over the WG and central SIK but overestimates connectivity in NIK. AgERA5 displays elevated degrees even in regions of weak observed coherence, indicating an overall tendency to over-smooth rainfall. In contrast, IMERG overestimates short-range coherence and suppress long-range connectivity, with degree distributions peaking at higher values than observed. These results demonstrate that network diagnostics expose structural differences among precipitation products that are not evident from conventional metrics, particularly in regions of complex terrain and strong monsoon forcing.

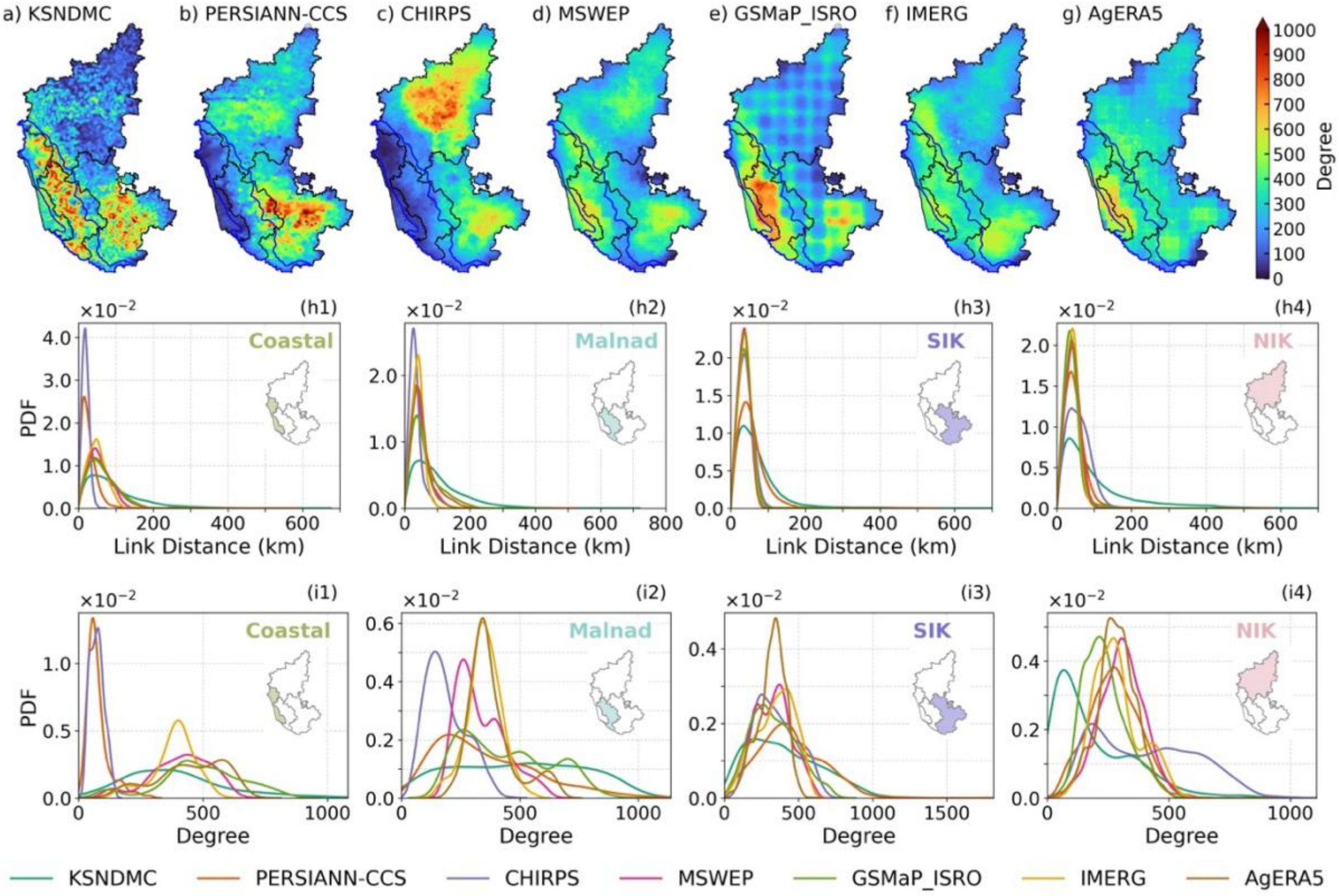

**Figure 2.** Spatial distribution of the degree (a–g) in the rainfall correlation network during the monsoon (JJAS) season constructed from daily mean rainfall for the period 2011–2022 (2011–2020 for MSWEP); (h1–h4) probability density function (PDF) of the link distance; (i1–i4) PDF of the degree.

The Salton cosine similarity (Figure 3) provides a measure of how well the gridded rainfall datasets reproduce the spatial connectivity structure of daily rainfall observed in gauge network. Higher values (close to 1) indicate that the dataset captures not only local rainfall characteristics but also the broader coherence pattern across the region. The analysis reveals that PERSIANN-CCS and CHIRPS exhibit low agreement with the observed dataset over the WG, indicating limited capability of these products to capture the terrain-driven rainfall variability in this complex region. In the NIK region, cosine similarity is more or less heterogeneous, with some areas showing low similarity and others moderate, reflecting the localized and patchy nature of rainfall in this region and this feature is consistent across datasets. Across SIK, most datasets show moderate to high similarity with the observed data, highlighting their relatively good ability to capture large-scale rainfall patterns in this less topographically complex region. However, AgERA5 show only moderate similarity in SIK, possibly due to its reliance on reanalysis inputs and smoothing, which may dilute finer-scale rainfall features. In the coastal region, MSWEP, GSMaP_ISRO, IMERG, and AgERA5 display similar spatial patterns of similarity—high cosine similarity in the southern coastal zone and moderate similarity in the northern part, likely driven by differences in orographic and mesoscale influences along the coastline. Among all datasets, GSMaP_ISRO shows the highest similarity with observations in the Malnad region, highlighting its advantage in capturing orographically influenced rainfall.

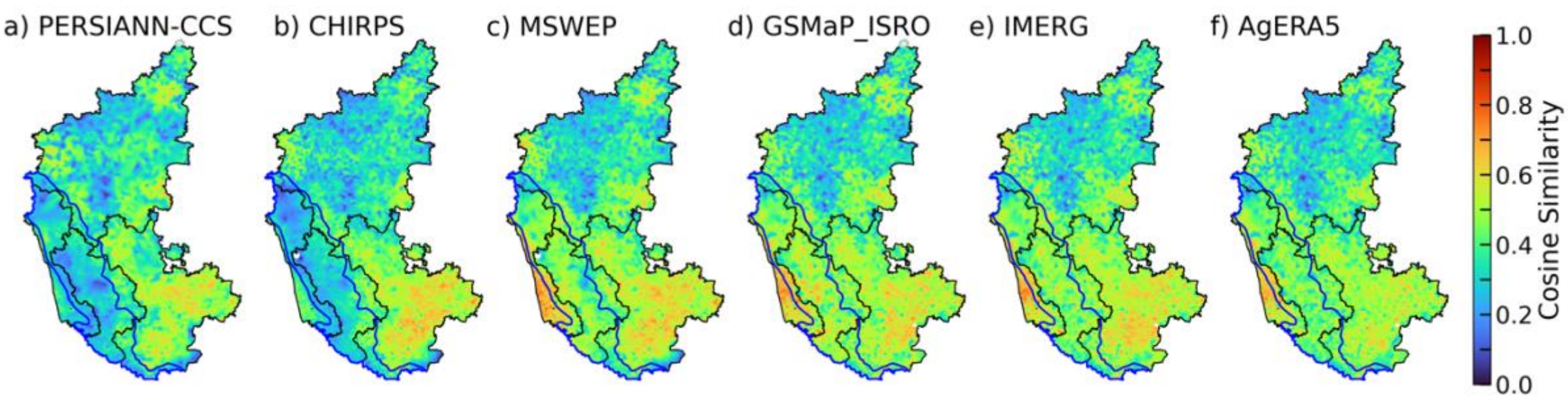

**Figure 3.** Spatial distribution of Salton cosine similarity between datasets and observed rain-gauge rainfall.

### 3.3. Extreme Rainfall Analysis

The spatial distribution of the degree, which represents the number of connections each station have with others during the synchronization of EREs is shown in Figure 4. A higher degree indicates that a station experiences rainfall extremes in synchrony with many other locations, suggesting the influence of large-scale atmospheric systems. In contrast, lower degree values point to more isolated stations, where EREs are likely driven by localized mesoscale processes, such as thunderstorms or orographic convection. The observed rainfall network shows strong synchronization between the coastal region and parts of central SIK and NIK regions, highlighting the broad spatial co-occurrence of extreme events in these areas. Meanwhile, the Malnad region displays relatively low degree values, implying that extreme events here tend to be more localized—likely a result of topographically induced rainfall confined to the windward slopes of the WG. Similarly, eastern parts of the SIK and NIK regions also show low degrees, reinforcing the interpretation that convective systems in these regions are more isolated and locally generated. Among the datasets, GSMaP_ISRO closely reproduces the observed degree distribution, particularly capturing the spatial pattern of synchronization between the coastal and inland regions, thereby reaffirming its consistency with ground-based observations. IMERG shows a broadly similar pattern, albeit with slightly higher degree values. In contrast, CHIRPS exhibits a different behavior—indicating higher degrees over the Malnad region and suggesting widespread synchronization between coastal extremes and those in the SIK and NIK regions, which may be an overrepresentation. PERSIANN-CCS, MSWEP, and AgERA5 display widespread high degrees across much of the state, implying exaggerated spatial co-occurrence of EREs and likely reflecting their tendency to smooth out localized features due to coarser resolution or interpolation-based bias correction methods.

The consistent link distance PDFs (Figure 4 h1–h4) across datasets suggest a broad agreement in the spatial scale over which extreme rainfall events co-occur. The similarity indicates that all products are capturing the spatial footprint of monsoon extremes reasonably. In contrast, the differing degree PDFs (Figure 4 i1–i4) reveal variation in the underlying topological structure of the extreme rainfall synchronization networks. These differences imply that the datasets vary in the number of such co-occurring events they detect likely due to differences in event detection thresholds, or spatial smoothing techniques. Thus, while the spatial extent of co-occurrence is comparably represented, the frequency and density of synchronization differ across datasets, influencing the inferred network connectivity of EREs.

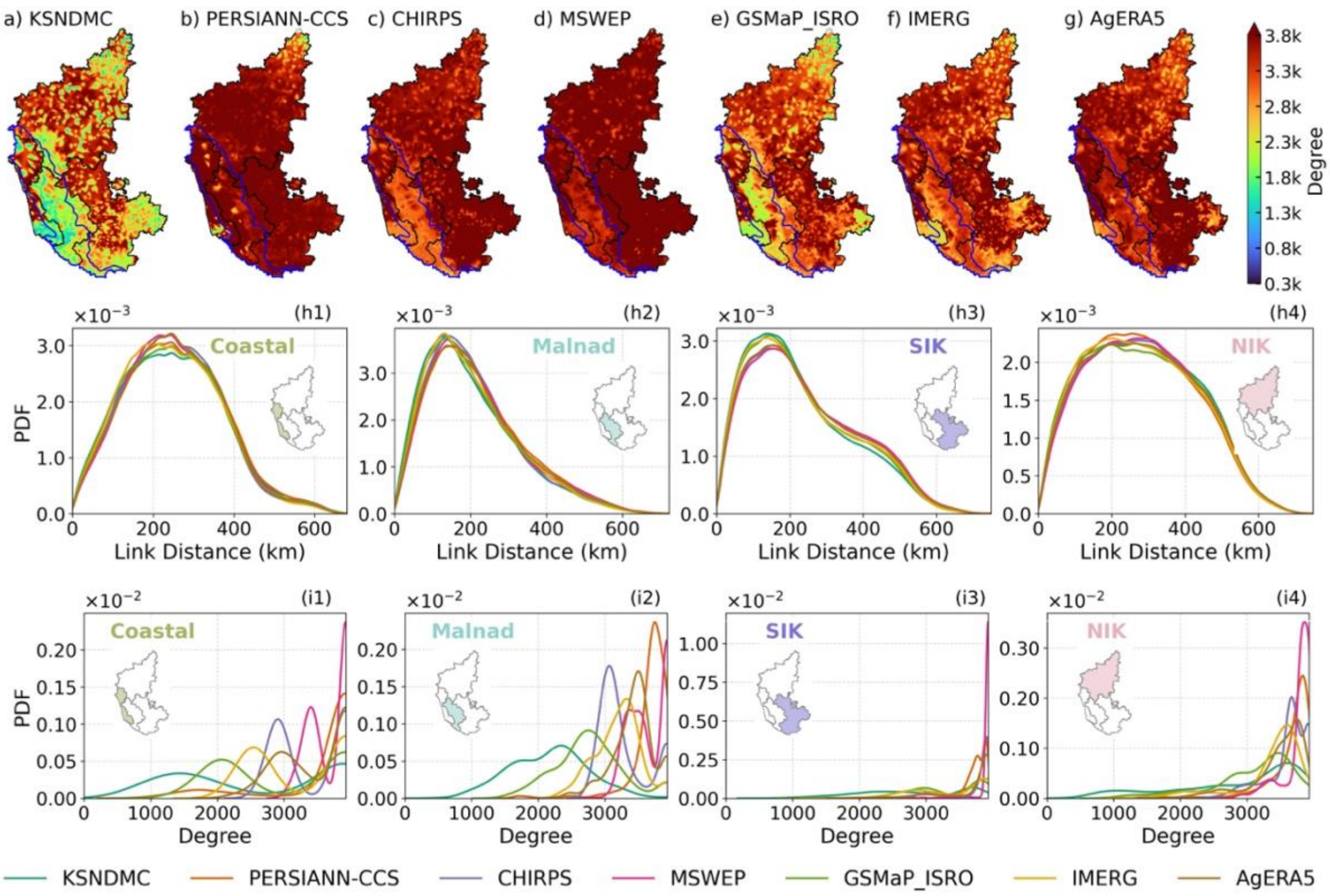

**Figure 4.** Spatial distribution of the degree values representing synchronization of extreme rainfall events during the summer monsoon (JJAS) for the period 2011–2022 (2011–2020 for MSWEP) in summer monsoon rainfall (JJAS) season; (h1–h4) PDF of the link distance; (i1–i4) PDF of the degree.

## 4. Conclusion

Advances in satellite remote sensing have led to the development of numerous global and regional precipitation products that provide continuous spatial and temporal coverage. However, these datasets remain subject to uncertainties arising from differences in retrieval algorithms and limitations in available ground-based observations. In this study, complex network metrics were employed to evaluate the spatiotemporal organization of gridded precipitation products relative to dense rain-gauge observations, providing complementary insight beyond conventional scalar error metrics. The evaluation was conducted over Karnataka, a southern Indian state characterized by complex topography and a pronounced west–east rainfall gradient, demonstrating the utility of network-based approaches for diagnosing structural precipitation errors in gridded precipitation products.

Over the interior regions of Karnataka, most gridded precipitation products show limited differentiation when assessed using conventional metrics, with the exception of PERSIANN-CCS. However, network-based analysis of the spatial correlation structure reveals clear and systematic differences among the datasets relative to the observed organization. While all products exhibit regional signatures linked to their respective retrieval approaches, GSMaP_ISRO shows the closest overall agreement with the observed rainfall structure across Karnataka. The rainfall correlation networks identify the Malnad region and central SIK as highly connected zones that act as hubs anchoring synoptic and mesoscale rainfall systems. In contrast, other products display region dependent deviations from the observed structure, with PERSIANN-CCS shows improved alignment over SIK and enhanced coherence along the eastern Malnad region, consistent with link-distance statistics, whereas CHIRPS exhibits over-coherence in the locally connected NIK. MSWEP, IMERG, and AgERA5 broadly reproduce the observed spatial organization but with generally weaker connectivity, except for localized enhancements over NIK. Among all datasets, GSMaP_ISRO consistently captures the observed structural characteristics across regions, with reduced coherence over central SIK likely reflecting the influence of bias correction using IMD gridded observations. Cosine similarity analysis further indicates weaker agreement of PERSIANN-CCS and CHIRPS with rain-gauge observations over complex terrain, although all products exhibit approximately 70% agreement over SIK and mixed performance across NIK. Finally, extreme-event synchronization highlights strong localization over the WG and southern coastal Karnataka, in contrast to longer-range synchronization over interior regions, a behavior consistently reflected in the link-distance distributions across datasets.

Overall, these results demonstrate that network-based diagnostics provide a robust framework for identifying structural differences in precipitation products that are not readily captured by traditional evaluation metrics. Such approaches offer a physically meaningful pathway for improving the assessment of satellite precipitation products over complex terrain and may contribute to more reliable characterization of regional rainfall variability and extremes.

## Acknowledgements

A.B., P.V., and R.I.S. are grateful for funding and support provided by the Institute of Eminence (IoE) project under the grant number SP22231222CPETWOCTSHOC.

**Authors Contribution**

**Ajay Bankar:** Conceptualization, Data curation, Formal Analysis, Visualization, Writing – original draft; **Praveenkumar Venkatesan:** Formal Analysis, Visualization, Writing – review and editing; **Rakesh V:** Formal Analysis, Writing – review and editing, Supervision; **Gaurav Chopra:** Supervision Writing – review and editing; **R I Sujith:** Conceptualization, Project Administration, Supervision, Writing – review and editing.

**Code Availability**

The codes used in this study will be provided upon reasonable request.